\documentclass[%
 reprint,
 amsmath,amssymb,
 aps,
]{revtex4-2}

\usepackage{graphicx}
\usepackage{dcolumn}
\usepackage{bm}
\usepackage{xcolor}
\usepackage{amssymb}
\usepackage{amsmath}
\usepackage{float} 
\usepackage{hyperref}
\usepackage{xcolor}
\usepackage{afterpage}
\usepackage[export]{adjustbox}
\usepackage{adjustbox}
\usepackage{caption}
\usepackage{subcaption}

\begin{document}
\preprint{APS/123-QED}

\title{Scaling and shape of financial returns distributions modeled as conditionally independent random variables}

\author{Hernán Larralde}
\affiliation{Instituto de Ciencias Físicas - Universidad Nacional Autónoma de México, Cuernavaca, 62210, Morelos, México}

\author{Roberto Mota Navarro}
\email{motanavarro@gmail.com}
\affiliation{Instituto de Ciencias Físicas - Universidad Nacional Autónoma de México, Cuernavaca, 62210, Morelos, México}
\affiliation{Università degli Studi di Palermo, Palermo, 90133, Sicily, Italy}

\date{\today}

\begin{abstract}
We show that assuming that the  returns are independent when conditioned on the value of their variance (volatility), which itself varies in time randomly, then the distribution of returns is well described by the statistics of the sum of conditionally independent random variables.
In particular, we show that the distribution of returns can be cast in a simple scaling form, and that its functional form is directly related to the distribution of the volatilities. This approach explains the presence of power-law tails in the returns as a direct consequence of the presence of a power law tail in the distribution of volatilities. It also provides the form of the distribution of Bitcoin returns, which behaves as a stretched exponential, as a consequence of the fact that the Bitcoin volatilities distribution is also closely described by a stretched exponential. We test our predictions with data from the S\&P 500 index, Apple and Paramount stocks; and Bitcoin.

\end{abstract}

\maketitle

\section{Introduction}

In finance, stylized facts are simple statistical properties of empirical data -mostly related to prices, but also observed in trade volumes \cite{navarro2025empirical}- that are generally true for a wide variety of instruments, time periods, and markets. 
In \cite{chakraborti2011,cont2001}, Chakraborti and Cont list a set of common stylized statistical facts. Of these, the following pertain directly to this work:
Absence of linear autocorrelations (i.e. linear autocorrelations
of asset returns are negligible after a short period of time); heavy tails (the distribution of returns often displays a power-law tail, with exponent between 2 and 5) and volatility clustering (volatility displays a positive autocorrelation over several days, implying that high/low volatility events tend to cluster in time). 

Another common feature that asset return distributions share with a large range of systems, is that they present scaling behavior \cite{mantegna1995scaling}. This property might provide important insights into the nature and inner workings of financial markets, and guide the construction of models that may help in the understanding of these systems. As it turns out, a model consistent with these facts would be simply to assume that, \emph{given} a certain value of the volatility (defined as the standard deviation of the returns calculated over a period of a day), returns can be considered as independent, and the volatility that describes the fluctuations of the returns takes random values in time. 
Indeed, the main result of this paper is that, under this simple assumption, and given the fact that the returns are additive, the distribution of returns over time intervals that comprise a large number of transactions, is well described by the statistics of the sum of conditionally independent random variables \cite{Hernan2024}. In particular, we show that the distributions of returns over different time intervals obey a simple scaling law. We also show that the heavy (power law) tails of the return distribution, when they occur, are essentially a consequence of the volatility distribution. We also show that returns distributions better described by stretched exponentials, can also be obtained from this perspective for appropriate volatility distributions. 

If we denote the price of a given financial instrument at time $t$ as $P(t)$, then the logarithmic returns over a time interval $\tau$ are defined as follows:
\begin{equation*}
r_{\tau}(t) = \ln\left( \frac{P(t)}{P(t-\tau)} \right)
\end{equation*}
(we take $\tau$ to be one minute). The returns behave additively when compounded over $n$ intervals. That is, defining the return over $n$ time intervals as  
\begin{equation*}
r_{n\tau}(t) : = \ln \left( \frac{P(t)}{P(t-n\tau)} \right)
\end{equation*}
It follows that
\begin{align*} 
r_{n\tau}(t) &= \ln \left( \frac{P(t)}{P(t-\tau)} \frac{P(t-\tau)}{P(t-2\tau)} \ldots \frac{P(t-(n-1)\tau)}{t-n\tau} \right) \\
                &=  \sum_{j=0}^{n-1}r_\tau(t-j\tau)
\end{align*}
As mentioned above, these individual returns are considered to be independent random variables when conditioned on the value of the volatility that describes the fluctuations during the period of time over which they are sampled. These volatilities also fluctuate in time; however, the fact that returns present volatility clustering, justifies assuming that for short time scales, the volatility that describes the fluctuations of the individual returns $r_\tau(t)$ over the $n$ time steps to yield the compound return $r_{n\tau}(t)$ is essentially constant. This implies that when sampled over a long period of time, comprising epochs of different volatilities, the joint distribution of  successive individual returns can be described by
\begin{equation}
f_n\left( r_{\tau}(t_1),...,r_{\tau}(t_n)\right) = \int  \left\{\prod_{i=1}^{n} f(r_{\tau}(t_i)|\sigma)\right\} h(\sigma)d\sigma 
\end{equation}
where $h(\sigma)$ is the distribution of volatilities along the sample and $f(r_{\tau}|\sigma)$ denotes the probability distribution for the individual returns conditioned on a given value of the volatility. Note that this implies that the returns are \emph{not} independent, as the distribution is not equal to the product of the marginal densities. However, it is easy to see that if the mean $\langle r_\tau(t)|\sigma\rangle=0$, the \emph{linear} auto correlation function is zero for all lags greater than zero, in agreement with the stylized facts. 

Making the assumption of conditional independence allows us to use the results for rescaled sums of this type of variables developed in \cite{Hernan2024}.  Thus, the rescaled distribution of logarithmic returns over $n$ time steps $F_n(r_{n\tau})$,  should display the following functional form

\begin{equation}\label{eqn:scaledsum}
    F_n(z) \approx \frac{1}{\sqrt{2n\pi}}\int \frac{1}{\sigma}e^{- z^2/2n\sigma^2}h(\sigma)d\sigma
\end{equation} 
Where $h(\sigma)$ is the distribution of standard deviations, $n\gg 1$, and since we are considering time scales of up to a few hours, we have assumed that the average return is indeed negligible.

Our assumption that returns are independent variables when conditioned on the volatility implies that the central limit theorem is applicable over epochs of a given volatility. This not only explains the observation in \cite{warusawitharana2018} that a conditional normal distribution with time-varying volatility fits the data best, but, as we will demonstrate later, predicts the scaling behavior of the returns distributions as the interval over which the returns are calculated varies.

\section{Data and methodology}

We analyze logarithmic returns from time series of prices of the S\&P 500 index and two of its components: Apple Inc. (AAPL) and Paramount Pictures Corporation (PARA), the largest and one of the smallest companies on the list of S\&P 500 components, respectively, and also of Bitcoin (BTC). The data was bought from provider FirstRate Data (\url{https://firstratedata.com}). We chose AAPL and PARA to confirm that our predictions are not affected by market capitalization.

For all series, we computed the standard deviations of 1 min logarithmic returns over non-overlapping rolling windows with a length of 390min, which corresponds to the duration of a trading day, and then proceeded to estimate the distribution of those standard deviations. 

All series are sampled every 1 minute and span the years 2007 to 2022  (S\&P 500 and PARA), 2005 to 2022 (AAPL) and 2019 to 2024 (BTC). When logarithmic returns at lower sampling resolutions were necessary, we resampled the 1 minute prices at the desired frequency. 

If Eq.(\ref{eqn:scaledsum}) holds, then it is straight forward to see that it should have the following simple diffusive-like scaling form
\begin{equation}\label{eq:scale}
    F_{n}(r) \sim \frac{1}{\sqrt{n}}\mathcal{F}(\frac{r}{\sqrt{n}})  \qquad n\gg 1
\end{equation}
However, the scaling function $\mathcal{F}(z)$ is not necessarily a Gaussian, as it would be if the returns were independent random variables.
To obtain the functional forms of the scaled returns distributions predicted by Eq.(\ref{eqn:scaledsum}), we performed numerical fits of the tails of the standard deviation distributions $h(\sigma)$ and insert the resulting models into the expression in Eq.(\ref{eqn:scaledsum}).

\section{Results}
\subsection{S\&P 500, AAPL and PARA}

\begin{figure*}
    \includegraphics[width=\textwidth]{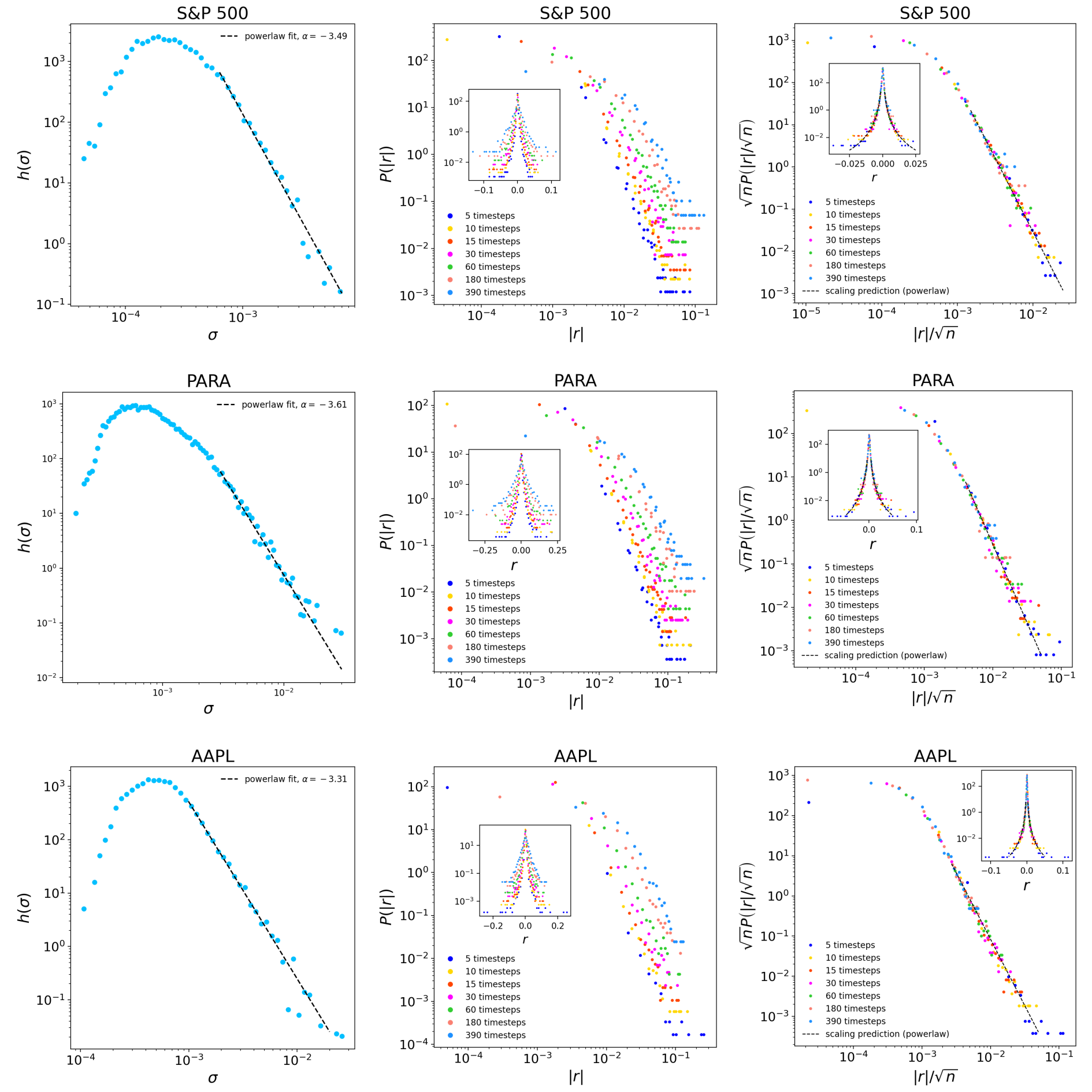}
    \caption{Unscaled returns distributions (1st column), volatility distributions (2nd column) and scaled returns distributions (3rd column) for the S\&P 500 index (1st row), and the stocks Apple Inc. AAPL (2nd row) and Paramount Global PARA (3rd row).  The unscaled returns distributions (first column), calculated over different time scales are shown in different colors. In all cases, the distributions corresponding to each time scale have different widths. Since the distributions are reasonably symmetric, we show the distributions of the absolute value of returns in order to visualize both tails in the same plot in a doubly logarithmic scale. The insets show the distributions in semi-log scale.
    The volatilities (second column) are computed as the standard deviations over non-overlapping windows containing 180 samples of returns at a scale of 1min. The asymptotic behavior of the right tails are well described by power laws, fitted using the algorithm described in \cite{clauset2009} via the Python package \texttt{powerlaw}\cite{alstott2014}. The exponents and the ranges over which each power law is a good fit are computed systematically by the package. Finally, the distributions of absolute returns after scaling (third column), along with the behavior predicted by equation \ref{eqn:scaledsum} when we substitute $h(\sigma)$ with the power law extracted from the volatility distributions. We observe that the distributions collapse onto a single curve, and that the tails of the distribution are well described by the expected power laws.}
\label{fig:STOCKS}
\end{figure*}

In the first column of figure \ref{fig:STOCKS} we show the distributions of the absolute value of the returns  for the S\&P 500, AAPL and PARA at different sampling rates before scaling. We chose AAPL and PARA because AAPL presently represents the largest component of S\&P 500, while PARA represents one of the smallest.

We see that returns at different time scales display distributions with widths that increase as the number of time steps grows. The insets show show semi-log plots of the complete distributions. As expected, at these time scales the distributions are reasonably symmetric.

The distributions of volatilities are shown in the second column of figure \ref{fig:STOCKS} and it is apparent that a power law does a good job of modeling the behavior of the tail. This power law behavior for the distribution of volatilities for certain assets was reported some time ago \cite{liu1999,gopikrishnanscaling}.

From equation (\ref{eqn:scaledsum}) and the results of the power law fit on the volatility distribution, we expect the rescaled returns distributions to display data collapse and to have a power law behavior at the tail, with the same exponent as the one found for the volatilities. This is indeed the behavior we observe, as shown in the third column of figure \ref{fig:STOCKS}.

\subsection{Bitcoin}

\begin{figure*}
    \includegraphics[width=\textwidth]{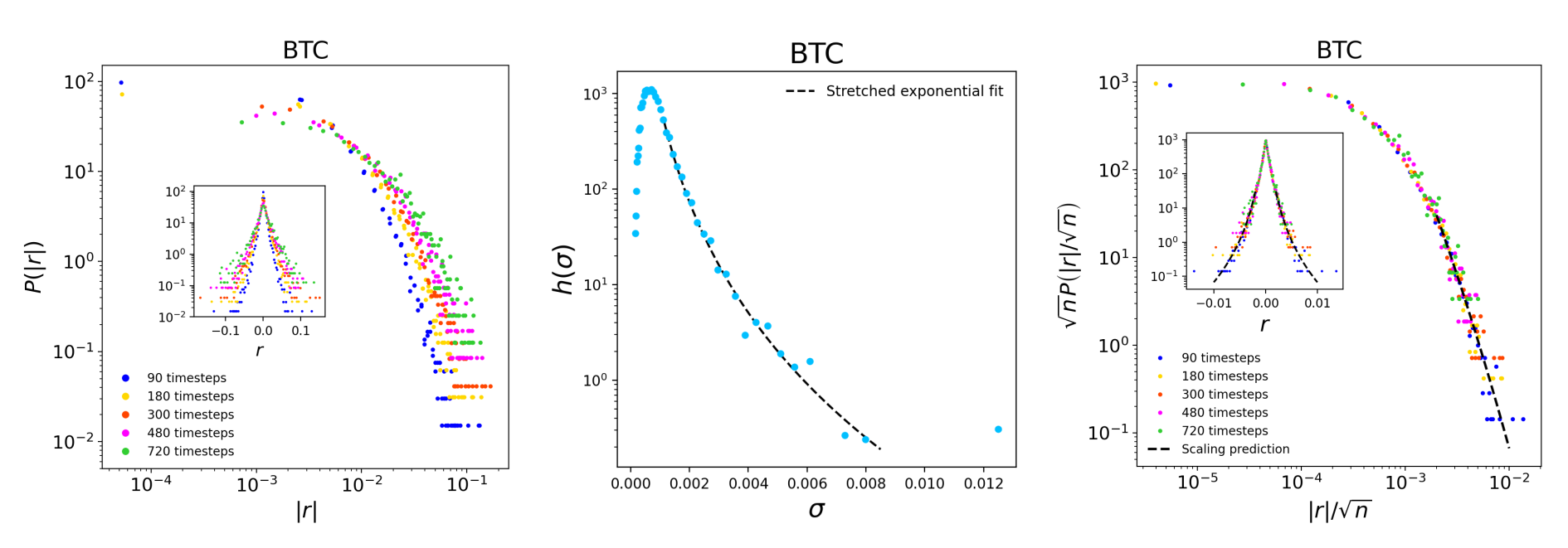}
    \caption{Absolute returns distribution for BTC before scaling (first plot), the volatility distribution (second plot) and the rescaled returns distribution (third plot) along with the fit for the right tail. With regards to the unscaled returns, BTC presents different characteristics from the stocks discussed previously. Nevertheless, different time scales in the measurement of returns also result in distributions with different widths. Since the distributions again are reasonably symmetric, we show the distribution of the absolute value of returns in order to visualize both tails in the same doubly logarithmic plot. The inset shows the densities in semi-log scale. As in the previous cases, the volatilities are computed as the standard deviations over non-overlapping windows containing 180 samples of returns at a scale of 1min. The asymptotic behavior of most of the right tail is well described by a stretched exponential (fitted via least squares), though there are certainly outliers that do not follow the stretched exponential law, but have little weight in the distribution. The third plot, with the absolute BTC returns after scaling shows how, again, all the distributions collapse onto a single curve. The dashed line corresponds to the eq. (\ref{eqn:I(x)}), obtained by substituting $h(\sigma)$ with the stretched exponential extracted from the volatility distribution into equation \ref{eqn:scaledsum}.}
    \label{fig:BTC}
\end{figure*}

We now show results for the Bitcoin/USD market. 
The distribution of unscaled BTC returns is shown in the first panel of figure \ref{fig:BTC}, where we confirm again that different sampling resolutions result in different distribution widths.
However, unlike what we observe in the S\&P 500 and its components, neither the returns nor the volatilities display power-law behavior in the tail of their respective distributions. This behavior has been observed already on previous studies (\cite{wkatorek2021} and references therein)

The distribution of volatilities is shown in figure \ref{fig:BTC}, we can see there that Bitcoin volatilities do not display the typical power-law decay in the tails of the previous stock returns, but rather, most of the tail can be fitted by a stretched exponential, that is $h(\sigma)=Ce^{-\lambda \sigma^{\beta}}$, though clearly there are outliers that are not described by the stretched exponential. However, these events have very little weight in the distribution.

By substituting $h(\sigma) \sim Ce^{-\lambda \sigma^{\beta}}$ into equation \ref{eqn:scaledsum} and performing a simple asymptotic approximation, we get that the distribution of Bitcoin returns at the tails should be described by the following expression:

\begin{equation}\label{eqn:I(x)}
  I(z;\lambda) \sim \frac{1}{z^{\beta/(2+\beta)}}e^{-\frac{\beta+2}{2\beta}\big(\lambda \beta z^\beta\big)^{2/(\beta+2)}}, 
\end{equation}
where $\lambda=61.38,\beta=0.1772$ are the parameters of the stretched exponential obtained by fitting the right tail of the volatility distribution. This is indeed the behavior we observe in the rescaled returns, as shown in figure \ref{fig:BTC}.

\section{Discussion} 

By assuming that the returns can be modeled as conditionally independent random variable, we are able to predict the scaling properties of the return distributions and connect the form of the scaling function with that of the distribution of volatilities. We test these predictions with the returns associated with the S\&P 500 index, as well as its largest and smallest (by market value) components. In all three cases we obtained a good data collapse and heavy tails, in agreement with the known stylized facts.

Our assumption, that returns are independent variables when conditioned on the volatility implies that the central limit theorem is applicable over epochs of a given volatility. This not only explains the observation in \cite{warusawitharana2018} that “conditional normal distribution with time-varying volatility fits the data best”, but predicts the scaling behavior of the returns distributions as the interval over which the returns are calculated varies.

On the other hand, unlike stocks, Bitcoin returns do not present power-law behavior at the tails of their distribution nor does the distribution of Bitcoin volatilities. However, most models found in the literature, having been developed to describe power-law behavior\cite{doyne2004}, are inadequate descriptors of the tails of distributions with different shapes, as is the case with Bitcoin. Thus, if we recur to one of the many models found in the literature capable of generating or describing parametrically\cite{gopikrishnan1998} power-law behavior, we would need to use another model for the particular case of Bitcoin returns. Instead, the assumption of independence conditional on volatility, provides a reasonably accurate description of the tails of the returns distribution for both power law tailed distributions, common in stocks and other financial instruments, as well as for Bitcoin, where the departure from power law behavior can be attributed to the unusual behavior of its volatility distribution.

In contrast to instrinsically parametric approches, like this work \cite{dedomenico2023} which compares popular models introduced in econophysics and finance to capture the heavy-tailed fluctuations of return distributions, our approach provides a simple unique connection between the parametric description of the returns, as a direct consequence of the parametric description of the statistics of the volatility.

We also want to stress out that our work does not \emph{explain} the origin of the heavy tail behavior of the distributions, we do not invoke any kind of universality as in \cite{gabaix2003} nor assign it to finite size effects as in \cite{alfi2009,alfi2009minimal}.

We are well aware that the predicted scaling form in eq. (\ref{eq:scale}) differs from the scaling found in \cite{mantegna1995scaling}. There, they propose that the distribution of \emph{arithmetic} returns has a scaling form $P(r,t)\sim t^{-1/a}\mathcal{F}(r/t^{1/a})$ with $a\approx 1.4$ and $\mathcal{F}(z)$ a Lévy stable distribution. While both data collapses are reasonably good, our scaling form is an unavoidable consequence of the assumption that returns are conditionally independent random variables.

\subsection{Of outliers and smaller time scales}

The stretched exponential fit of the BTC volatility distribution stops being an adequate description of the right tail at the most extreme values of the volatility $\sigma$. Still, although the outliers are not properly described by a stretched exponential, the asymptotic expression we obtain when considering only the portion of the tail modeled by it is a reasonably accurate model of the tails of rescaled returns distributions as seen in Figure (\ref{fig:BTC}). This is not surprising when we take into consideration that the outliers represent a very small portion of the events, accumulating a minuscule probability mass in relation to the probability mass accrued in the exponential part. 
Thus, even when we ignore the outliers and assume that a stretched exponential decay is the full description of the right tail, the asymptotic expansion (Eq. \ref{eqn:I(x)}) remains a good model of the rescaled returns. This is \emph{not} to say that the outliers are not important. Outliers reflect important events that, though rare, may have strong financial repercussions. 

As for the time scales, the predictions of the model lose accuracy when we go below the time scales shown in the results. For the case of stocks (S\&P 500, AAPL and PARA), this means going below 5 minutes. For the case of BTC, the loss of accuracy starts below 90 minutes. We believe that this is a consequence of the non-negligible autocorrelations in returns that are present at intraday frequencies, mainly as a consequence of the bid-ask bounce\cite{cont2001}. These correlations violate the assumption of conditional independence, so the loss of predictive power of the model is to be expected. Further exploration of these cases is a possible future development.

\bibliography{apssamp}

\end{document}